\documentclass[aps,pra,twocolumn]{revtex4}
\usepackage{graphicx}

\usepackage{bm}

\newcommand{\br}[1]{\langle #1|}
\newcommand{\ke}[1]{|#1\rangle}

\newcommand{\da}{^\dagger}
\newcommand{\pt}[1]{\left( #1 \right)}
\newcommand{\pq}[1]{\left[ #1 \right]}

\newcommand{\av}[1]{\left\langle #1 \right\rangle}
\newcommand{\al}[1]{^{(#1)}}

\begin{document}

\title{Ground state cooling in a bad cavity}
\author{Stefano Zippilli$^{1,2,3}$}
\author{Giovanna Morigi$^3$}
\author{Wolfgang P. Schleich$^1$}
\affiliation{
$^1$ Abteilung Quantenphysik, Universit\"at Ulm, 89069 Ulm, Germany\\
$^2$ ICFO - Institut de Ci\`encies Fot\`oniques, 08860 Castelldefels (Barcelona), Spain
\\
$^3$ Grup d'Optica, Departament de Fisica, Universitat Aut\`onoma de Barcelona, 08193 Bellaterra, Spain
}

\date{\today}
\begin{abstract}
We study the mechanical effects of light on an atom trapped in a harmonic potential when an atomic dipole transition is driven by a laser and it is strongly coupled to a mode of an optical resonator. We investigate the cooling dynamics in the bad cavity limit, focussing on the case in which the effective transition linewidth is smaller than the trap frequency, hence when sideband cooling could be implemented. We show that quantum correlations between the mechanical actions of laser and cavity field can lead to an enhancement of the cooling efficiency with respect to sideband cooling. Such interference effects are found when the resonator losses prevail over spontaneous decay and over the rates of the coherent processes characterizing the dynamics. 
\end{abstract}
\maketitle

\section{Introduction}

Sideband cooling of atoms in harmonic traps has been demonstrated to be a successful technique for preparing the atomic center-of-mass in states of high purity~\cite{Leibfried03,Eschner03}. This technique exploits the coupled dynamics of external and internal degrees of freedom due to a laser drive, in the regime in which the energy levels of the center-of-mass oscillator can be spectrally resolved~\cite{Stenholm86}. This regime can be achieved by properly choosing atomic species which offer a dipole or quadrupole transition whose linewidth fulfills this requirement. In absence of a suitable transition, the desired enhancement of the scattering processes leading to cooling can be achieved by appropriately coupling atomic levels, as in the case of Raman-sideband cooling~\cite{Leibfried03,Eschner03}, or by means of an optical resonator, thereby exploiting the modified structure of the electromagnetic field~\cite{Cirac95,Vuletic01,vanEnk,KimbleFORT03,Buschev04,Kuhn05,Zippilli05a,Zippilli05b,Domokos03}. 
It should be remarked that the discreteness of the spectrum of the center-of-mass motion, which is here a harmonic oscillator, 
may give rise to peculiar scattering properties, which are due to interference between the mechanical excitations induces by the mechanical effects of light, and may result in a critical enhancement of the rate of scattering into certain atomic levels~\cite{Cirac95,Zippilli05a}. As a consequence, parameter regimes can be encountered, where the cooling efficiency is appreciably enhanced. In particular, in~\cite{Zippilli05b} ground state cooling has been predicted in a broad parameter regime for atomic transitions coupled to good resonators, even when the atomic transition linewidth does not allow for spectrally resolving the excitations of the center-of-mass oscillator.

In this article we investigate how the dynamics of a trapped atom is modified by the presence of an optical resonator, when the resonator decay is the prevailing loss mechanism. We consider an atom confined in a bad resonator and driven transversally by a laser, in the regime in which the motion can be sideband cooled to the ground state in free space. We start from the equations presented in~\cite{Zippilli05a}, which we rederive using the resolvent formalism~\cite{Cohen}, and study the predicted dynamics. We find that interference between the mechanical effects of resonator and laser can appreciably enhance the ground state cooling efficiency, which can result larger than sideband cooling. Such dynamics are accessed when the resonator decay rate exceeds by several orders of magnitude the trap frequency, and exhibit a non-trivial dependence on the geometry of the setup. They could be observed in experimentally accessible parameter regimes. 

This article is organized as follows. In Sec.~\ref{Sec:Model} the model is introduced, and the basic equations for the
motion are derived. In Sec.~\ref{Sec:III} we discuss the dynamics of cooling. In Sec.~\ref{Sec:Conclusions} the conclusions are drawn. 

\section{Model}
\label{Sec:Model}

\begin{figure}[h]
\includegraphics[width=6.5cm]{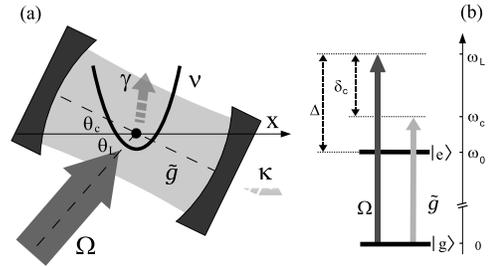}
\caption{(a) A mode of an optical resonator couples with strength $\tilde g$  to the atomic dipole, 
which is driven transversally by a laser at Rabi frequency $\Omega$. The atomic motion is confined by an external harmonic potential at frequency $\nu$. The system dissipates by spontaneous emission of the atomic excited state at rate $\gamma$
and by cavity decay at rate $\kappa$. (b) Internal dipole structure and comparison among the frequency of the laser, $\omega_L$, of the dipole transition, $\omega_0=\omega_L-\Delta$, and of the cavity mode $\omega_c=\omega_L-\delta_c$. Further parameters are defined in Sec.~\ref{Sec:Model}.} \label{Fig:model}
\end{figure}

We consider an atom of mass $M$, which is confined by a harmonic
potential of frequency $\nu$ inside an optical resonator.
The atomic dipole couples to a mode of the cavity field and to a laser, as shown in Fig.~\ref{Fig:model}(a). 
The atom internal degrees
of freedom, which are relevant to the dynamics, are the ground
state $\ke{g}$ and the excited state $\ke{e}$, constituting a
dipole transition at frequency $\omega_0$.
The cavity mode is at frequency $\omega_c$ and wave vectors ${\bf k_c}$, the laser is at frequency $\omega_L$ and wave vector ${\bf k_L}$. The internal structure and couplings are depicted in Fig.~\ref{Fig:model}(b), where $\Delta=\omega_L-\omega_0$ and $\delta_c=\omega_L-\omega_c$ denote the detunings of the laser from dipole and cavity, respectively. Dipole and cavity mode are coupled to the external modes of the electromagnetic field into which the atom can spontaneously emit and to which the cavity field decays by the finite transmittivity of the mirrors. 

In the following, we consider the quantum mechanical motion of the atomic center of mass. In particular, we restrict to the motion along the $x$-axis, and neglect the motion in the transverse plane assuming tight transversal confinement. This assumption simplifies the treatment, and at the same time allows to highlight the basic features of the dynamics. In particular, we treat systematically the dependence of the mechanical effects on the geometry of the setup, here given by the angles $\theta_c$ and $\theta_L$ of cavity and laser wavevectors with the motional axis, as shown in Fig.~\ref{Fig:model}(a), such that the wavevectors components along $x$ are ${\bf k_c}_x=k\cos\theta_c$ and ${\bf k_L}_x=k\cos\theta_L$. 

The Hamiltonian describing the quantum dynamics of atomic dipole, center of mass, and electromagnetic field (e.m.f.) modes, in the reference frame rotating at the laser frequency, is
\begin{eqnarray}
H_{\rm tot}=H_0+W
\end{eqnarray}
where $H_0$ describes the coherent dynamics in absence of
coupling between atom and e.m.-field, and reads
\begin{eqnarray}
\label{Hmec}
H_0=\hbar\nu b^{\dagger}b-\hbar\Delta\sigma\da\sigma-\hbar\delta_ca\da a+H_{\rm emf}
\end{eqnarray}
Here, $b$, $b^{\dagger}$ are annihilation and creation operators of a quantum of vibrational energy $\hbar\nu$ of the center-of-mass oscillator, $\sigma=\ke{g}\br{e}$ is the dipole lowering operator, and $\sigma\da$ is its
adjointy; $a$, $a\da$ are the annihilation and creation operators of a cavity photon. The hamiltonian term $H_{\rm emf}$ describes the oscillators corresponding to the external e.m.f.-modes,
\begin{eqnarray}
H_{\rm emf}=-\sum_j\hbar\delta_j a_j\da a_j-\sum_k\hbar\delta_k a_k\da a_k
\end{eqnarray}
where we label with subscript $j$ the modes which couple to the dipole and with $k$ the modes which couple to the cavity mode through the finite mirror transmission. Here, $a_j$, $a_j\da$, $a_k$, $a_k\da$ are the corresponding annihilation and creation operators, with $\delta_j=\omega_L-\omega_j$ and $\delta_k=\omega_L-\omega_k$ their detunings from the laser frequency.
The coupling between atom and electromagnetic field is described in the electric dipole approximation by operator $W$, which we decompose as
\begin{equation}
W=W_j+W_k+H_L+H_{\rm at-cav}
\end{equation}
Here, the terms
\begin{eqnarray}\label{C3:reservoir}
W_j&=&\hbar\sum_j g_j \sigma a_j\da \left[1-{\rm i}\eta\cos\theta_j(b+b^{\dagger})\right]+{\rm H.c.}\\
W_k&=&\hbar\sum_k  f_k (a_k\da a+a_k a\da)
\end{eqnarray}
describe the interaction of the external e.m.f-modes with the dipole and the cavity, respectively, $g_j$ and $f_k$ are the coupling strength with the dimension of a frequency, and $\eta=\sqrt{\hbar k^2/2M}$ is the Lamb-Dicke parameter, weighting the mechanical effects of photon recoil. In Eq.~(\ref{C3:reservoir}) we have used the Lamb-Dicke expansion in first order~\cite{Stenholm86}. The radiative coupling of the atomic dipole with laser and cavity mode is described by operators $H_L$ and $H_{\rm at-cav}$, respectively. Using the Lamb-Dicke expansion, we decompose the latter terms into $H_L=H_L^{(0)}+H_L^{(1)}$, and $H_{\rm at-cav}=H_{\rm at-cav}^{(0)}+H_{\rm at-cav}^{(1)}$, where the superscript indicates the order in the parameter $\eta$. The terms giving the coupling with the laser read
\begin{eqnarray}
&&H_L^{(0)}=\hbar\Omega\sigma\da+{\rm H.c.}\\
&&H_L^{(1)}={\rm i}\hbar\eta\cos\theta_L\Omega\sigma\da(b+b^{\dagger})+{\rm H.c.}
\label{HL}
\end{eqnarray}
where $\Omega$ is the Rabi frequency. Finally, the terms giving the coupling with the cavity mode read
\begin{eqnarray}
\label{Hat-cav}
&&H_{\rm at-cav}^{(0)}=\hbar g \cos(\phi)a\da\sigma+{\rm H.c.}\\
&&H_{\rm at-cav}^{(1)}=-\hbar \eta \cos\theta_c g
\sin(\phi)a\da\sigma(b+b^{\dagger})+{\rm H.c.}
\end{eqnarray}
where $g$ is the cavity-mode vacuum Rabi frequency and $\phi$ is a phase which accounts for the position of the trap center in the mode spatial function. For later convenience, we denote the atom--cavity coupling strength at the trap center by
\begin{eqnarray*}
\tilde g=g\cos\phi
\end{eqnarray*}
and the coefficients, scaling the mechanical effects of laser and cavity, by
\begin{eqnarray*}
\varphi_L&=&\cos\theta_L\\
\varphi_c&=&\cos\theta_c\tan\phi
\end{eqnarray*}
through which the dependence on the geometry of the setup enters the problem.

\subsection{Scattering rates}

We now evaluate the rates of the scattering processes, which lead
to a change of the vibrational excitation of the center-of-mass
oscillator. We consider the lowest relevant order in the Lamb
Dicke parameter $\eta$. Moreover, we assume that the atom is
weakly driven by the laser, and take thus the Rabi frequency
$\Omega$ as a small parameter. We consider the transitions from the
initial state
\begin{eqnarray}\label{in}
\ke{\rm i}&=&\ke{g,0_c,n;0_j,0_k}\nonumber\\
\end{eqnarray}
at energy $E_i$, to the final states
\begin{eqnarray}\label{fin}
\ke{{\rm f}_{j,k}}&=&\ke{g,0_c,n\pm 1;1_{j,k}}
\end{eqnarray}
at energy $E_{f_{j,k}}$, where $\ke{g,0_c}$ indicates  the atomic internal
ground state and the cavity field vacuum state, $\ke{n}$ and
$\ke{n\pm 1}$ are the initial and final states of the harmonic
motion, as we consider only processes which change the motional state. 
The state $\ke{0_j,0_k}$ represents the external e.m.f.-vacuum state, 
the state $\ke{1_j}=\ke{1_j,0_k}$ one photon in one of the modes $j$ due to atomic emission, and 
the state $\ke{1_k}=\ke{0_j,1_k}$ one photon in one of the modes $k$ due to cavity decay.

The  transition amplitude from the initial state~(\ref{in}) to the final state~(\ref{fin}), beloging to a continuum spectrum, is the element of the scattering matrix
\begin{eqnarray}\label{C3:Scat}
S_{\rm f- i}\al{j,k}=\delta_{if}-2i\pi\delta^{(T)}\pq{E_{\rm i}-E_{f_{j,k}}}T_{\rm f- i}\al{j,k}
\end{eqnarray}
where $T_{\rm f- i}\al{j,k}$ is the element of the transition matrix, which at first order in $\Omega$ and $\eta$ reads
\begin{eqnarray}\label{T}
T_{\rm f-i}\al{j,k}
&=&\br{{\rm f}_{j,k}}W_{j,k}\frac{1}{E_{\rm i}-H_{\rm eff}}H_{L}^{(1)}\ke{\rm i}\\
& &+\br{{\rm f}_{j,k}}W_{j,k}\frac{1}{E_{\rm i}-H_{\rm eff}}H_{\rm at-cav}^{(1)}\frac{1}{E_{\rm i}-H_{\rm eff}}H_{L}^{(0)}\ke{\rm i}
\nonumber\end{eqnarray}
Here,
\begin{eqnarray}\label{Heff}
H_{\rm eff}=H_0+H_{\rm at-cav}^0-{\rm i}\hbar\frac{\kappa}{2}a\da a-{\rm i}\hbar\frac{\gamma}{2}\sigma\da\sigma.
\end{eqnarray}
is the effective Hamiltonian at zero order in $\Omega$ and $\eta$, and $\gamma$, $\kappa$ are the spontaneous decay and the cavity decay rates, respectively. They are given by 
$\gamma=2\pi|g_j(\omega_0)|^2\rho_j(\omega_0)$ and $\kappa=
2\pi|f_k(\omega_0)|^2\rho_k(\omega_c)$, with $\rho_{j,k}(\omega)$ density of states of the e.m.f.-modes $j,k$ at frequency $\omega$. 

The total transition rate into the states of the continuum $|{\rm f}_{j,k}\rangle$, leading to a change of the vibrational excitation by one quantum, is found from the time derivative of the squared modulus of $S_{\rm f- i}\al{j,k}$, after taking the sum over the final states $|{\rm f}_{j}\rangle$ and $|{\rm f}_{k}\rangle$. 
Eventually the transition rates read
\begin{eqnarray}\label{C3:transitionrates}
\Gamma_{n\to n\pm 1}=\Gamma_{n\to n\pm 1}^{\gamma}+\Gamma_{n\to n\pm 1}^{\kappa}
\end{eqnarray}
where $\Gamma_{n\to n\pm 1}^{\gamma}$ and $\Gamma_{n\to n\pm 1}^{\kappa}$ are the scattering rates for the processes in which the vibrational excitation is changed by spontaneous emission and by cavity decay, respectively, 
\begin{eqnarray}
&&\Gamma_{n\to n\pm 1}^{\gamma}=\gamma\eta^2\xi_\pm\pt{\alpha|T_S^\pm|^2+|\varphi_LT_L^{\gamma,\pm}+\varphi_cT_c^{\gamma,\pm}|^2}\nonumber\\
&&\label{C3:transitionrates:gamma}\\
&&\Gamma_{n\to n\pm 1}^{\kappa}=\kappa\eta^2\xi_\pm|\varphi_LT_L^{\kappa,\pm}+\varphi_cT_c^{\kappa,\pm}|^2
\label{C3:transitionrates:kappa}
\end{eqnarray}
where
\begin{eqnarray}
\label{C3:transitions}
{ T}_S^\pm&=&\Omega\frac{\delta_c+{\rm i}\kappa/2}{f(0)}\\
\label{C2:TLgamma}
{T}_L^{\gamma,\pm }&=&{\rm i}\Omega\frac{(\delta_c\mp\nu+{\rm i}\kappa/2)}
{f(\mp\nu)}\\
\label{C2:TLkappa}
{T}_L^{\kappa,\pm }&=&{\rm i}\Omega\frac{\tilde g}{f(\mp\nu)}\\
{T}_c^{\gamma,\pm }&=&-\Omega\frac{\tilde g^2(2\delta_c\mp\nu+{\rm i}\kappa
)}{f(0)f(\mp\nu)}  \\
{T}_c^{\kappa,\pm}&=&-\Omega\frac{\tilde
g\pq{(\Delta\mp\nu+{\rm i}\gamma/2) (\delta_c+{\rm
i}\kappa/2)+\tilde g^2}}{f(0)f(\mp\nu)}\label{C3:transitions:F}
\end{eqnarray}
with
\begin{eqnarray}
f(x)=(x+\delta_c+{\rm i}\kappa/2)(x+\Delta+{\rm i}\gamma/2)-\tilde g^2
\end{eqnarray}
Here, $\xi_+=n+1$, $\xi_-=n$, and
$$\alpha=\int_{-1}^1 {\rm d}\cos\theta_j\cos^2\theta_j {\cal N}(\cos\theta_j),$$
gives the angular dispersion of the atom momentum due to the
spontaneous emission of photons. These expressions agree with the ones found using a density matrix formalism in~\cite{Zippilli05a,Zippilli05b}. 

From rates~(\ref{C3:transitionrates}) the dynamics of the center-of-mass motion can be inferred when coherences between different motional number states are negligible. This assumption requires $\nu\gg\eta \Omega|\varphi_L|,\eta |\tilde{g}\varphi_c|$, and it is fulfilled in the parameter regime in which the rates have been derived. We can hence construct a rate equation for the occupation propability $p_n$ of the number state $\ke{n}$, 
\begin{eqnarray}\label{rateEq}
\frac{d}{dt}p_n&=&-\left(\Gamma_{n\to n+1}+\Gamma_{n\to n-1}\right)p_n\\
               & &+\Gamma_{n+1\to n}p_{n+1}+\Gamma_{n-1\to n}p_{n-1}\nonumber
\label{RateEq:pn}
\end{eqnarray}
These equations can be generalized to describe the dynamics of the center-of-mass in three dimensions, as they 
have been derived for any geometry of the setup. They
acquire the well-known form, usually encountered in the literature of laser cooling of trapped ions~\cite{Stenholm86}, when
writing
\begin{eqnarray}
\Gamma_{n\to n+1}&\equiv&\eta^2(n+1)A_+\\
\Gamma_{n\to n-1}&\equiv&\eta^2 n A_-,
\end{eqnarray}
with $A_{\pm}$ the so-called heating and coolig rates. In this manuscript we will characterize the cooling efficiency by the steady number state occupation $\langle n\rangle_{\rm St}=\sum_n np_n^{\rm St}$, where $\dot{p}_n^{\rm St}=0$ are the stationary occupation probabilities, and by the rate $W$ at which it is reached. The expectation value $\langle n\rangle_{\rm St}$ takes the simple form
\begin{eqnarray}
\label{n:steady} \av{n}_{\rm St}=\frac{A_+}{A_- - A_+}
\end{eqnarray}
for $A_->A_+$, and the cooling rate $W$ is 
\begin{eqnarray}
\label{Cool:Rate} W=\eta^2 (A_- - A_+)
\end{eqnarray}
In the following, we discuss the dependence of these quantities on the cavity and laser parameters in the regime where the prevailing loss mechanism is cavity decay, hence for $\kappa\gg\gamma$, and search for the optimal parameters leading to ground state cooling, $\langle n\rangle_{\rm St}\ll 1$. Moreover, we focus onto the regime where $\kappa\gg\nu$, and more specifically $\kappa\gg\nu\gg\gamma$.

\section{Ground state cooling in the bad cavity limit}
\label{Sec:III}

In this section we investigate cooling of the atom to its motional ground state
in the limit in which the rate of cavity decay $\kappa$ is the largest parameter in the system dynamics and the rate of spontaneous emission $\gamma$ is very small. In particular, we focus on the limit $\gamma\ll\nu\ll\kappa$. 
Let us remark that for these parameters ground state cooling can be already achieved in free space, as in the Lamb-Dicke regime and for $\gamma\ll\nu$ the basic conditions for efficient sideband cooling are fulfilled~\cite{Leibfried03,Eschner03,Stenholm86}. Hence, in this section we aim at identifying regimes, where the efficiency of sideband cooling can be rised by the presence of the resonator. Moreover, we search for novel dynamics, which can basically differ from sideband cooling, and still lead to an enhanced ground state occupation.

\subsection{Basic processes in the bad-cavity limit}

\begin{figure}[!t]
\includegraphics[width=8cm]{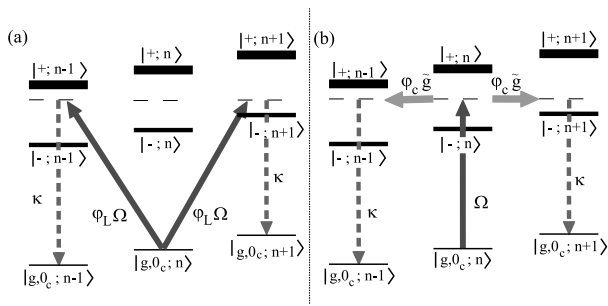}
\caption{Dominant scattering processes leading to a change of the
vibrational number by one phonon in the bad cavity limit $\gamma\ll\nu\ll\kappa$. The states $|g,0_c;n\rangle$,
$\ke{\pm;n}$ are the cavity-atom dressed states at phonon number
$n$. These processes  describe scattering of a laser photon
by cavity decay.
Here, (a) correspond to  ${T}_L^{\kappa,\pm}$ and  (b)  to ${T}_c^{\kappa,\pm}$, see text.
}
\label{fig:schemi}
\end{figure}

For $\kappa\gg\gamma$  the main processes contributing to Eq.~(\ref{C3:transitionrates}) describe scattering of a laser photon by cavity decay. Therefore in most cases heating and cooling rates are basically due to photon scattering by cavity losses,
$\Gamma_{n\to n\pm 1}\approx \Gamma_{n\to n\pm 1}^{\kappa},$ 
and $\Gamma_{n\to n\pm 1}^{\gamma}$ are in general small corrections. 
Let us now discuss in detail the dynamics described by the two terms adding up coherently in the transition rates~(\ref{C3:transitionrates:kappa}). They describe processes
where the motion is changed by mechanical coupling to the laser (${T}_L^{ \kappa,\pm }$) and to the
cavity (${T}_c^{\kappa,\pm }$) field. The process described by~${T}_L^{\kappa,\pm }$ scales with the geometric factor $\varphi_L$, which accounts for the recoil due to absorption of a laser photon. The transition
amplitude~${T}_c^{ \kappa,\pm }$ describes the mechanical coupling due to the resonator. It thus scales
with the geometric factor $\varphi_c$ which accounts for the recoil due to interaction with the cavity mode. Since the final
state of the two scattering processes is the same, state $\ke{\rm f_k}$, they interfere. 

These processes can be graphically represented considering the dressed states $\{\ke{\pm;n}\}$ as intermediate states of the scattering process, where 
\begin{eqnarray}
&&|+;n\rangle=\sin\vartheta |g,1_c;n\rangle + \cos\vartheta |e,0_c;n\rangle
\\
&&|-;n\rangle=\cos\vartheta |g,1_c;n\rangle -\sin\vartheta |e,0_c;n\rangle
\end{eqnarray}
with $\tan\vartheta=\tilde g/(-\Delta_c/2+\sqrt{\tilde g^2+\Delta_c^2/4})$ and
$\Delta_c$ the detuning between cavity mode and atom. Fig.~\ref{fig:schemi} represents the term ${T}_L^{ \kappa,\pm }$ in terms of transitions between $|g,0_c,n\rangle$ and these states, showing that the scattering rate is the coherent sum of six transition amplitudes, weighted by the geometrical factors $\varphi_L$ and $\varphi_c$. Hence, the dynamics are in general non-trivial, and may depend critically on the geometry of the setup. When the splitting between the dressed states is the largest parameter, the coupling to one of the dressed states is negligible. Hence, the scattering rate reduces to the sum of two transition amplitudes, and one recovers the result reported in~\cite{Cirac95}. In~\cite{Zippilli05b} it has been shown, that the result of~\cite{Cirac95} is  a particular limit of Eq.~(\ref{C3:transitionrates:kappa}).

\subsection{Efficiency of ground state cooling}

\begin{figure}[!t]
\includegraphics[width=7.5cm]{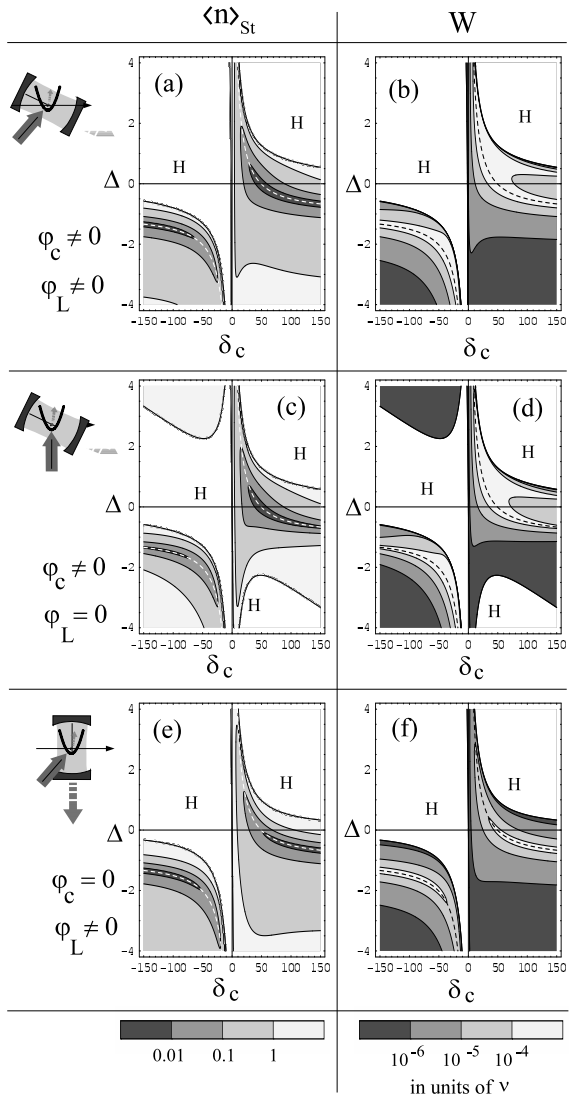}
\caption{Contour plots of the average phonon number at steady state $\av{n}_{\rm St}$ (left column)  and
corresponding cooling rate $W$ (right column) as a function of $\delta_c$ and $\Delta$ (in units of $\nu$) for $\gamma\ll\nu\ll\kappa$ and three possible geometries:
in the first row $\theta_L=\theta_c=\pi/4$: the mechanical effects of both cavity and laser contribute to the
dynamics. In the second row $\theta_L=\pi/2$ and $\theta_c=\pi/4$: the mechanical effects are solely due to the cavity. In the last row $\theta_L=\pi/4$ and $\theta_c=\pi/2$: the mechanical effects are solely due to the laser. In the contour plots the heating regions are not coded and explicitly indicated by the label H. The dashed lines indicate the curve $\delta_{\rm opt}(\Delta)$, Eq.~(\ref{deltaopt}). The parameters are $\eta=0.1$, $\phi= \pi/4$, $\Omega=0.03\nu$, $\tilde g=7\nu$, $\gamma=0.1\nu$, $\kappa=10\nu$.}
\label{fig:Bad}
\end{figure}

In this section we plot the results obtained from our analytical equations~(\ref{RateEq:pn}) with the rates~(\ref{C3:transitionrates})-(\ref{C3:transitionrates:kappa}). A comparison of their predictions with numerical simulations, using the quantum Monte-Carlo wavefunction method, has been presented in~\cite{Zippilli05b}, where a good agreement has been found in the regime of validity of the equations.

\begin{figure}[!t]
\includegraphics[width=8cm]{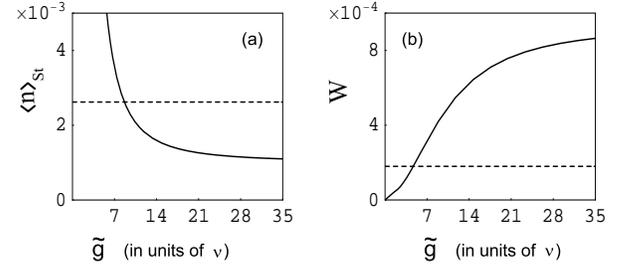}
\caption{Average phonon number at steady state $\av{n}_{\rm St}$ and cooling rate $W$, in units of $\nu$, as a function of the atom--cavity vacuum--Rabi coupling $\tilde g$ when the mechanical effects are due both to laser and cavity field ($\theta_L=\pi/4$ and $\theta_c=\pi/4$). The solid lines are evaluated for $\Delta=0$ and at the cavity detuning satisfying Eq.~(\ref{deltaopt}), $\delta_c=\delta_{\rm opt}(0)$. The dashed lines correspond to the standard sideband cooling limit, when $\Delta=-\nu$ and there is no coupling to the cavity mode. The other parameters are $\eta=0.1$, $\phi= \pi/4$, $\kappa=10\nu$, $\gamma=0.1\nu$, and $\Omega=0.03\nu$. }
\label{fig:Bad-g}
\end{figure}

Figure~\ref{fig:Bad} displays the average phonon number at steady state $\av{n}_{St}$ and the corresponding cooling rate $W$
as a function of $\delta_c$ and $\Delta$, for $\gamma\ll\nu\ll\kappa$ and in the strong coupling regime, $\tilde g^2/\gamma\kappa\gg1$, for different geometries, corresponding to the cases where the mechanical effects of the resonator and of the laser contribute with different weights to the cooling dynamics. The dashed curve in the contour plots represents the function~\cite{Zippilli05a,Zippilli05b} 
\begin{eqnarray}\label{deltaopt}
\delta_{\rm opt}(\Delta)\equiv \frac{\tilde g^2+\gamma\kappa/4}{\Delta+\nu}-\nu.
\end{eqnarray}
for which $A_-$ is maximized. This corresponds to choose the parameters in order to set the red sideband transition at a resonance of the atom--cavity system~\cite{Zippilli05b}. As it is visible from the contour plots, high cooling efficiencies are obtained in the parameters region about this curve. Nevertheless, the cooling efficiency in certain parameter regimes depend critically on the geometry - and thus on whether the mechanical effects are due to the cavity or to the laser. In
Fig.~\ref{fig:Bad}(c)-(d), for instance, we see a region of efficient cooling for $\delta_c<0$ and $\Delta>0$, which shrinks substantially in Fig.~\ref{fig:Bad}(e)-(f), where the cavity wave vector is perpendicular to the motion, and hence the mechanical effects originate solely from the coupling with the laser. In particular, in Fig.~\ref{fig:Bad}(c) one sees that low temperatures are achieved for a broad interval of values of the detunings about $\Delta=0$ and $\delta_c>0$. From comparison with Fig.~\ref{fig:Bad}(e)-(f) it is clear, that these dynamics are particularly sustained by the mechanical effects of the cavity mode. This parameter regime is analyzed in Fig.~\ref{fig:Bad-g}, where we compare the cooling efficiency obtained by taking $\Delta=0$, with standard sideband cooling. Here, it is visible that the coupling to the resonator appreciably enhance the efficiency, such that lower temperatures and larger cooling rates than with sideband cooling are predicted. 

\begin{figure}[!t]
\includegraphics[width=6cm]{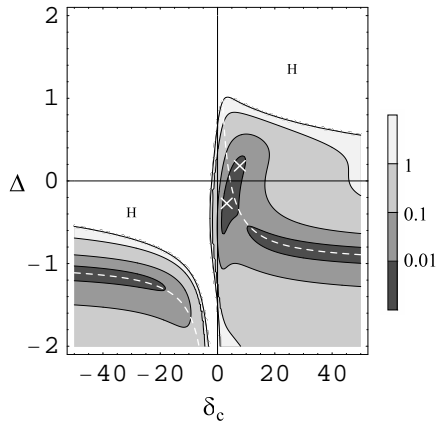}
\caption{Contour plot of the average phonon number at steady state $\av{n}_{\rm St}$ as a function of $\delta_c$ and $\Delta$ (in units of $\nu$) for $\gamma\ll\nu\ll\kappa$ and $\theta_L=\theta_c=\pi/4$. Here, $\eta=0.1$, $\phi= \pi/4$, $\kappa=10\nu$ (as in Fig.~\ref{fig:Bad}(a)) but $\gamma=0.01\nu$ and $\tilde g=2.3\nu$. The crosses indicates the points corresponding to $\{\delta_c^{\pm},\Delta^{\pm}\}$. Here, $\delta_c^+=7.8\nu$, $\Delta^+=0.2\nu$, $\delta_c^-=3.2\nu$, $\Delta^-=-0.3\nu$. }
\label{fig:Bad2}
\end{figure}

Interference phenomena, leading to suppression of transitions, can be found in particular parameter regimes. In general, one can identify the parameters, which lead to the vanishing of the blue sideband transition, $\Gamma_{n\to n+1}^{\kappa}=0$. They are identified by solving 
\begin{equation}
\label{Condition:1}
\varphi_LT_L^{\kappa +}+\varphi_cT_L^{\kappa +}=0
\end{equation}
The set of solutions of Eq.~(\ref{Condition:1}) includes the result discussed in~\cite{Cirac95} as a special limit, which is found when $|\delta_c|$ is the largest parameter. In general, Eq.~(\ref{Condition:1}) is solved for two different pairs of values $\{\delta_c^{\pm},\Delta^{\pm}\}$, provided that $\varphi\tilde{g}^2/\kappa\nu>1$ for $\varphi>0$ or $\tilde{g}^2/\kappa\nu|\varphi|>1$ for $\varphi<0$, with $\varphi=\varphi_L/\varphi_c$. This interference could thus be encountered in two very different physical regimes, either when $\tilde{g}^2/\kappa\nu\gg 1$ or $\tilde{g}^2/\kappa\nu\ll 1$, depending on the value of $\varphi$, and thus on the geometry of the setup. This interference effect is not visible in Fig.~\ref{fig:Bad}. In fact, it is in general washed away at finite values of $\gamma$, as in the parameter regime satisfying Eq.~(\ref{Condition:1}) there is appreciable scattering by spontaneous emission, $\Gamma_{n\to n+1}^{\gamma +}$. Figure~\ref{fig:Bad2} shows the average number of excitations at steady state as a function of $\Delta$ and $\delta_c$ at sufficiently small $\gamma$, in the parameter regime where the enhancement of the cooling efficiency due to this interference effect is visible. This corresponds to the region that stretches about two points, corresponding to $\delta_c^{\pm}$ and $\Delta^{\pm}$. The region is broad, showing that the cooling efficiency is robust against fluctuations around the values of these parameters. It 
corresponds to dynamics where the heating rate is suppressed due to destructive interference between the mechanical effects of laser and cavity. 

\section{Conclusions}
\label{Sec:Conclusions}

We have investigated the cooling dynamics of atoms confined in bad resonators by an external harmonic potential. This study focusses onto ground state cooling, and it considers the situation in which the linewidth of the atomic transition is smaller than the trap frequency, therefore in the regime in which sideband cooling can be implemented in free space. We have identified novel parameter regimes in which efficient ground state cooling is achieved, whose dynamics are sustained by the presence of the resonator and whose efficiency is appreciably better than sideband cooling. This occurs when the cavity decay rate exceeds by orders of magnitude the trap frequency. 

This work complements the investigations reported in~\cite{Zippilli05a,Zippilli05b}, which focussed onto the good cavity limit, and considers situations which could be observed in present experimental setups~\cite{KimbleFORT03,Guthorlein01,Keller04,Mundt02,Sauer03,Kuhn05,Kuhn05b}. In general, these results contribute to a further understanding of the complex dynamics of the mechanical effects of optical resonators on atoms, whose wealth of phenomena could be eventually exploited for implementing coherent control of this kind of systems.

\acknowledgments

The authors acknowledge support from the IST-network QGATES, the Integrated Project SCALA (Contract No. 015714), 
and the Scientific Exchange Programme Germany-Spain (HA2005-0001 and D/05/50582). G.M. is supported by the spanish Ministerio de Educacion y Ciencias (Ramon-y-Cajal and FIS2005-08257-C02-01).

\end{document}